\documentstyle[12pt]{article}

\begin{document}
\setcounter{page}{0}
\begin{flushright}
{CERN--TH.7350/94}\\
\end{flushright}
\medskip
\begin{center}
{\bf $N\bar{N}$ ANNIHILATION AT THE OPEN CHARM THRESHOLD}
\end{center}
\vskip1cm
\begin{center}
{\bf Boris Kerbikov$^{(a)}$
 and Dmitri Kharzeev\footnote{On leave of absence from Moscow State
University, Moscow, Russia}$^{(b,c)}$}
\end{center}
\medskip
\begin{center}
{\it (a) Institute for Theoretical and Experimental Physics\\ Moscow 117259,
Russia}\\
\vskip0.4cm
{\it (b) Theory Division\\CERN\\ CH-1211, Geneva 23, Switzerland}
\end{center}
\vskip0.4cm
\begin{center}
{\it (c) Physics Department, University of Bielefeld\\ 33615
Bielefeld, Germany}\\
\end{center}
\vskip0.7cm
\begin{abstract}
We discuss the $p\bar{p}$ annihilation into a pair of charmed $D$ mesons
not far from the threshold of the reaction. An interesting
interference pattern in the differential cross section is predicted
to arise from the
existence of
both $t$-channel charmed-baryon exchange and $s$-channel
$\Psi^{''}$ charmonium resonance amplitudes.
We argue that the experimental study of this process would provide
new and valuable information about the unknown $p\bar{p}\Psi^{''}$ and
$\Lambda_c N D$ couplings. These are important to know to solve
the long-standing puzzle of the heavy flavour contents of baryons and
to test the semi-local duality.
\end{abstract}
\vskip3cm
CERN--TH.7350/94\\
July 1994
\newpage

There exist many meson states which couple to the $N\bar{N}$ system.
These states then appear as the poles in the amplitudes of the $N\bar{N}$
interaction and this, in principle, can lead to observable effects.
However the common situation in the light quark sector is that
the widths of the resonances are large; they therefore overlap and
interfere.
The necessity to account, in addition, for an interference with
the (generally unknown) $t$-channel exchange amplitudes further
complicates the
situation, making a clean and detailed analysis
an extremely difficult task.

In this sense a rather uncommon situation is encountered in the
$p\bar{p}$ annihilation into a pair of $D$ mesons close to the threshold.
The $s$-channel resonance here is the $\Psi^{''}(3770)$ charmonium
state with a relatively small width of $\Gamma\simeq 23.6$ MeV. The nearest
charmonia are too far and narrow to cause any appreciable interference
effects, so the pole corresponding to $\Psi^{''}$ can, to a good
accuracy, be  considered as isolated.
In addition, there exist of course non-resonant $t$-channel amplitudes,
corresponding to the exchange of the states with the quantum numbers
of charmed baryon(s). The relation between the strengths of the resonant
and non-resonant amplitudes depends on two couplings,
$N\bar{N}\Psi^{''}$ and $N_c N D$. Almost nothing is known
at present about the values of the corresponding coupling constants,
apart from a few rather controversial predictions \cite{Rob93}-\cite{Vol93}.
%\cite{Kha92}, \cite{Bie92},
% \cite{Iof93}, \cite{Vol93}.
A possible admixture of the light quark--antiquark pairs in the
wave function of the $\Psi^{''}$ can further complicate the
situation, giving rise to a non-perturbative contribution
to the $p\bar{p}\rightarrow \Psi^{''}$ amplitude.
Another interesting issue concerning the $N_c N D$ coupling is that
the knowledge of its value is desirable to solve the problem of
 the intrinsic heavy flavour contents
of the baryons [6,7].
%\cite{Bro88},\cite{Ell89}.
The arguments listed above indicate that the study
of the $p\bar{p}\rightarrow D\bar{D}$ reaction is interesting, but at
the same time leave seemingly little chance to make a reliable prediction
for the relative importance of $s$- and $t$-channel exchange amplitudes.
\vskip0.3cm

There exists, however, an attractive theoretical scheme which, when
applied to our situation, gives a definite answer -- the (semi-local)
duality [8--11].
%\cite{Ven}, \cite{Sch}, \cite{Gat}, \cite{Dol}.
Namely, the semi-local duality implies that the averages (over a
finite--energy segment) of
 cross--sections
corresponding to the sum of all possible amplitudes in $s$- and $t$-channels,
respectively, should be approximately the same.

We therefore can expect the appearance of an interesting interference pattern
in the vicinity of the $\Psi^{''}$ resonance, arising from the competing
$s$- and $t$-channel amplitudes of comparable size.
To see in more detail how this pattern can appear, we shall try to calculate
directly the cross-section of the $p\bar{p}\rightarrow D\bar{D}$ process.
\vskip0.3cm

The $s$-channel resonant amplitude is given by
$$ T^{(s)} = - {{A(\Psi^{''}\rightarrow p\bar{p}) A(\Psi^{''}
\rightarrow D\bar{D}})
\over {s - M_R^2 + iM_R \Gamma_R}}, \eqno(1)$$
where $M_R$ and $\Gamma_R$ are the resonance mass and width and
$A(\Psi^{''}\rightarrow X)$ stands for the resonance decay amplitude.

For what concerns the $t$-channel amplitude, we shall model it by the
$\Lambda_c$ charmed hyperon exchange. We choose the effective  Lagrangean
of $\Lambda_c N D$ interaction as
$$ \L_{int} = i g (\bar{\Psi} \gamma_5 \Psi) \Phi, \eqno(2)$$
where $g\equiv g_{\Lambda_c N D}$ is the corresponding coupling constant,
$\Phi$ is the field of the $D$ meson, and $\Psi$ is the Dirac spinor
of the baryon.
The interaction (2) induces the $t$-channel exchange amplitude
of the form
$$ T^{(t)} = {{i g^2} \over {t - M^2}}\ \bar{v} (\hat{q} + \Delta) u,
\eqno(3)$$
where $M$ and $m$ are the charmed hyperon and nucleon masses, respectively,
$\Delta = M - m$, and $q$ is the c.m.s. momentum of the $D$ meson
(in what follows we shall keep the notations and
metrics conventions of Bjorken and Drell \cite{Bjo}).
\vskip0.3cm

The initial- and final-state interactions must also be taken into
account. If $S_{p\bar{p}}$ and $S_{D\bar{D}}$ are the $S$--matrix
elements describing the interaction in the initial and final channels,
respectively, then the total transition $S$ matrix element can be
represented as \cite{Tab85}
$$ S_{if} = \sqrt{S_{p\bar{p}}} \tilde{S_{if}} \sqrt{S_{D\bar{D}}}, \eqno(4) $$
where $\tilde{S_{if}}$ is the transition matrix element induced by the
amplitudes (1) and (3).

A fair description of the $p\bar{p}$ interaction in the energy range
of interest for us is given by the so-called Frahn--Venter (spin-dependent)
model \cite{Fra}. It has been successfully applied to describe the $p\bar{p}$
scattering \cite{Dau} and initial--state interaction in the
$p\bar{p}\rightarrow \Lambda\bar{\Lambda}$ reaction \cite{Ker}.
We thus feel free to introduce the initial-state interaction in
exactly the same way as it was done in \cite{Ker}. The main effect
of the initial--state interaction is to damp the cross section by
a factor of $\sim 10^{-2}-10^{-3}$, depending on the partial waves
involved.

The properties of the $D\bar{D}$ interaction
are essentially unknown; it nevertheless seems natural that it is much
weaker than the strongly absorptive $p\bar{p}$ interaction.
In what follows we shall put $S_{D\bar{D}}=1,$
thus neglecting the final--state scattering.
\vskip0.3cm

We now have to consider the amplitudes of $\Psi^{''}$ decay entering into the
resonant amplitude (1). A natural (though not necessarily true --
see \cite{Lip}) assumption is that the OZI-allowed $D\bar{D}$ mode
dominates the $\Psi^{''}$ decay; this fixes the $A(\Psi^{''}\rightarrow
D\bar{D})$ amplitude by the relation $\Gamma(\Psi^{''}\rightarrow
D\bar{D}) \simeq \Gamma_R$.

The partial width $\Gamma(\Psi^{''}\rightarrow p\bar{p})$ is presently unknown;
we thus have to make use of available theoretical predictions [1]-[5].
%\cite{Kha92}, \cite{Bie92}, \cite{Rob93}, \cite{Iof93}, \cite{Vol93}.
They are quite contradictory: while Refs. \cite{Kha92} and  \cite{Bie92}
predict the value of
$\Gamma(\Psi^{''}\rightarrow p\bar{p})$ to be in the range of $3-8$ eV,
and \cite{Rob93} -- the value of $\simeq 40$ eV,
Ref.[4] claims a value as high as $500\pm150$ eV. Keeping in mind
the controversial character of these theoretical predictions, we
shall use a rather conservative value of $\simeq7\ eV$, arising from
the appropriate scaling of the $J/\Psi$ and $\Psi^{'}$ partial decay
widths to the $p\bar{p}$ channel \cite{Kha92}, \cite{Bie92}.
The corresponding resonant part
of the $p\bar{p} \rightarrow D\bar{D}$ cross section at the peak
(without taking into account any of the interference effects) can therefore
be estimated as
$$ \sigma^{res}(p\bar{p}\rightarrow D\bar{D}) \simeq {{12 \pi}
\over{M_R^2-4m^2}} \ {{\Gamma(\Psi^{''}\rightarrow p\bar{p})} \over \Gamma_R }
\sim 0.5\ {\rm{nb}} \eqno(5)$$
\vskip0.3cm

The last ingredient that has to be fixed is the value of the
$\Lambda_c p D$ coupling constant.
To the best of our knowledge, it has not been evaluated previously.
Our first $SU(4)$--motivated guess for it is
$$ {{g^2_{\Lambda_c p D}} \over {4\pi}} \simeq {{g^2_{\Lambda p K}} \over
{4\pi}} \simeq 13.9 \pm 2.6, \eqno(6)$$
where the last value is taken from Martin's analysis \cite{Mar81} of
kaon--nucleon scattering data using the forward dispersion relation
techniques. The values extracted by other authors from the analyses
of kaon--nucleon scattering \cite{Ant}, kaon photoproduction \cite{Ade},
hyperon--nucleon potential \cite{Mae}, and the $p\bar{p}\rightarrow \Lambda
\bar{\Lambda}$ reaction \cite{Tim} are consistent with this result.

The evaluation of the amplitude (3) is more conveniently performed
in the helicity basis. The two helicity amplitudes corresponding to
the values of the total helicity $\Lambda=\lambda_p - \lambda_{\bar{p}}$
equal to $\Lambda = 0$ and $\Lambda =1$, respectively, are
$$ T^{(t)}_{++} = {{ i g^2} \over {t - M^2}} \left( \Delta {p \over m} +
q\ \cos \theta \right) ; \eqno(7a) $$
$$ T^{(t)}_{+-} = {{ i g^2} \over {t - M^2}} \left( { - E \over m} q\
\sin \theta \right) ; \eqno(7b) $$
where $p$ and $E=\sqrt{p^2 + m^2}$ are the momentum and energy of
the (anti-) proton in the c.m.s., and $\theta$ is the angle
between the directions of the incoming antiproton and outcoming
$D^{-}$ or $D^{0}$ meson.

The machinery of the evaluation of the amplitudes (7) is standard [17],
and includes the decomposition of the helicity amplitudes into
the components with definite total momentum. It is further convenient
to expand the denominator
$${1 \over {(t-M^2)}} = - { 1 \over {2pq}} {1 \over {z - \cos \theta}}, $$
where
$$ z= {1 \over {pq}} \left[ {1 \over 2} (M^2 - m^2 - M_D^2) + {s \over 4}
\right] $$
($M_D$ is the $D$-meson mass, and $s$ is the c.m.s. energy squared),
using the Heine formula
$$ (z - \cos \theta)^{-1} = \sum_{m=0}^{\infty} (2m+1) P_m(\cos \theta)
Q_m(z),$$
with $P_m$ being the Legendre polynomials and $Q_m$ the Legendre functions
of the second kind.
\vskip0.3cm

We now have to correct for the initial--state interaction.
The $S$ matrix elements of the $p\bar{p}$ scattering
that enter formula (4) are parametrized \cite{Dau}, \cite{Ker}
in the $\{LSJ\}$ basis,
where $L$ and $S$ refer to the orbital momentum and total spin in the
$p\bar{p}$ system.
To implement the initial-state interaction one has therefore
to make the transformation of the partial-wave helicity amplitudes
into $\{LSJ\}$ ones, which can be done easily:
$$ T^{J}_{L=J-1} = \sqrt{2 \over {2J+1}} \left( \sqrt{J} T_{++}^J
+ \sqrt{J+1} T_{+-}^J \right); \eqno(8a)$$
$$ T^{J}_{L=J+1} = \sqrt{2 \over {2J+1}} \left(- \sqrt{J+1} T_{++}^J
+ \sqrt{J} T_{+-}^J \right). \eqno(8b)$$
The parity conservation allows only the transitions $(J=L \pm 1, S=1)
\rightarrow (l = J)$, where $l$ is the orbital momentum in the
$D\bar{D}$ system.
\vskip0.3cm

Adding the $s-$channel resonant amplitude (1) to the
$t-$channel one,
we find ourselves in a position to calculate the cross section
\footnote{Summing the $s-$ and $t-$channel amplitudes seems to contradict
the idea of duality and could lead to double counting in computing
the total reaction cross section. Nevertheless, it is clear that an
interference
between these amplitudes should manifest itself in the differential
cross section in the vicinity
of a resonance.}.
The use of the $SU(4)$-motivated value
(6) for the $\Lambda_c p D$ coupling constant leads to the $t-$channel
contribution which is three orders of magnitude larger than that of the
$s-$channel $\Psi^{''}$ resonance, thus leading to a result that is
in strong contradiction with the semi-local duality.
We do expect however that due to the large
difference in the masses
of charmed and strange quarks the $SU(4)$ is severely broken,
and the value (6) is largely overestimated.

To perform an estimate of the cross section, we can however proceed
in a different way, choosing the magnitude of the
$\Lambda_c p D$ coupling so that the energy-averaged contributions
of the $s-$ and $t-$channel amplitudes become equal, according to
the semi-local duality \cite{Ven}, \cite{Sch}. The resulting value
is about
$$ {{g^2_{\Lambda_c p D}} \over {4\pi}} \simeq 0.5, \eqno(9)$$
which is some 30 times less than the $SU(4)$ motivated value (6).
A very interesting interference pattern then appears in the differential
cross section.

As a typical example of our results we show in Fig. 1 the differential cross
section of the $p\bar{p} \rightarrow D \bar{D}$ reaction at
$\sqrt{s}=3.79$ GeV,
i.e. $20$ MeV above the $\Psi^{''}(3.77)$ mass.
One can clearly see how the $\cos \theta$ distribution
of the $D$ mesons arising from the decay of the $\Psi^{''}$ resonance is
distorted by the asymmetric $t-$channel-exchange amplitude.
The magnitude of the cross section that we find is of the same order
as found by {\cite{Kroll.89}} and {\cite{Kaidalov.94}} in a different
framework.

The appearance of the interference pattern is extremely sensitive to the
relative strength of the $p\bar{p}\Psi^{''}$ and
$\Lambda_c N D$ couplings; its very existence requires these
couplings to
be of the same order of magnitude, in agreement with the semi-local
duality.

To summarize: we have calculated the differential cross section of the
$p\bar{p} \rightarrow D\bar{D}$ reaction close to its threshold.
We have found an interesting structure in the angular distributions of
$D$ mesons, arising from the interference between the resonant $s-$channel
and non-resonant $t-$channel amplitudes.
The very appearance of this interference pattern requires, however,
the strength of the two amplitudes to be approximately the same, as
is required by semi-local duality. Moreover, we find that the
shape of the angular distributions and the value of the total cross section
altogether allow one to determine the $p\bar{p}\Psi^{''}$ and
$\Lambda_c N D$ couplings unknown at present. The latter
are important to know since they are possibly driven by the non-perturbative
effects sensitive to the
heavy-flavour contents of baryons and/or to the light quark--antiquark
component of the $\Psi^{''}$.

An additional interest in the production of charmed mesons in $N\bar{N}$
annihilation not far from the threshold arises from the possible existence
\cite{Kon} of the exotic diquark--antidiquark
resonances with hidden charm, which would manifest themselves as additional
$s-$channel resonances.

We therefore consider the experimental measurement of the $p\bar{p} \rightarrow
D\bar{D}$ reaction as very desirable.

\vskip0.5cm

{\it{Acknowledgements:}}
We wish to thank R. Cester, B.L. Ioffe, A.B. Kaidalov, L.A. Kondratyuk,
R. Landua, M. Mandelkern, M.G. Sapozhnikov, J. Schultz, F. Tabakin and
P.E. Volkovitsky for useful and stimulating discussions.

D.K. acknowledges financial support from the German Research Ministry
(BMFT) under Contract 06 BI 721.

\newpage
\newpage


\begin{thebibliography}{999}

\bibitem{Rob93} R.W. Robinett and L. Weinkauf, {\it{Phys. Lett.}} {\bf{B271}}
(1991) 231;

\bibitem{Kha92} D.E. Kharzeev, {\it{Sov. J. Nucl. Phys.}} {\bf{55}} (1992)
1353;

\bibitem{Bie92} J.K. Bienlein, {\it{in: Proc. of the ``SuperLEAR" Workshop,
eds.C. Amsler and D. Urner}} (IOP Conf. Series No.124, London, 1992) 87;

\bibitem{Iof93} B.L. Ioffe, {\it{Phys. Rev.}} {\bf{D47}} (1993) 340;

\bibitem{Vol93} P.E. Volkovitsky, {\it{Phys. Lett.}} {\bf{B308}} (1993) 100;

\bibitem{Bro88} S.J. Brodsky, P. Hoyer, C. Peterson and N. Sakai,
{\it{Phys. Lett.}} {\bf{B93}} (1980) 451;

\bibitem{Ell89} J. Ellis, E. Gabathuler and M. Karliner, {\it{Phys. Lett.}}
{\bf{B217}} (1989) 173;

\bibitem{Ven} G.C. Rossi and G. Veneziano, {\it{Phys. Rep.}} {\bf{63}}
(1980) 149;

\bibitem{Sch} A.A. Logunov, L.D. Soloviev and A.N. Tavkhelidze,
{\it{Phys. Lett.}} {\bf{B24}} (1967) 181;

\bibitem{Gat} R. Gatto, {\it{Phys. Rev. Lett.}} {\bf{18}} (1967) 803;

\bibitem{Dol} R. Dolen, D. Horn, and C. Schmid, {\it{Phys. Rev.}} {\bf{166}}
(1968) 1768;

\bibitem{Bjo} J.D. Bjorken and S. Drell, {\it{``Relativistic quantum fields"}}
(McGraw - Hill, New York, 1965);

\bibitem{Tab85} F. Tabakin and R.A. Eisenstein, {\it{Phys. Rev.}} {\bf{C31}}
(1985) 1857.

\bibitem{Fra} W.E. Frahn and R.E. Venter, {\it{Ann. Phys.}} {\bf{27}} (1964)
135, 385, 401;

\bibitem{Dau} C. Daum et al., {\it{Nucl. Phys.}} {\bf{B6}} (1968) 617;

\bibitem{Ker} G. Schneider--Neureither et al., {\it{Z. Phys.}} {\bf{A344}}
(1993) 317;

\bibitem{Lip} H. Lipkin, {\it{Phys. Lett.}} {\bf{B179}} (1986) 278;

\bibitem{Mar81} A.D. Martin, {\it{Nucl. Phys.}} {\bf{B179}} (1981) 33;

\bibitem{Ant} J. Antolin, {\it{Z. Phys.}} {\bf{C31}} (1986) 417;

\bibitem{Ade} R.A. Adelseck and B. Saghai, {\it{Phys. Rev.}} {\bf{C42}}
(1990) 108;

\bibitem{Mae} P.M.M. Maessen, Th.A. Rijken, and J.J. de Swart,
{\it{Phys. Rev.}} {\bf{C40}} (1989) 2226;

\bibitem{Tim} R.G.E. Timmermans, Th.A. Rijken, and J.J. de Swart,
{\it{Phys. Lett.}} {\bf{B257}} (1991) 227;

\bibitem{Ber} L.D. Landau, V.B. Berestetsky, E.M. Lifshitz, and
L.P. Pitaevsky, {\it{``Relativistic quantum field theory"}} (Pergamon,
Oxford, 1974);

\bibitem{Mou} B. Moussalam, {\it{Nucl. Phys.}} {\bf{A429}} (1984) 429;

\bibitem{Kroll.89} P. Kroll, B. Quadder and W. Schweiger,
{\it{Nucl. Phys.}} {\bf{B316}} (1989) 373;

\bibitem{Kaidalov.94} A.B. Kaidalov and P.E. Volkovitsky, to be published;

\bibitem{Kon} D.E. Kharzeev and L.A. Kondratyuk, {\it{in: Proc. of
the "Diquarks" Workshop, eds. M. Anselmino and E. Predazzi,}}
(World Scientific, 1994).
\end{thebibliography}
